\documentclass[a4paper,11pt]{article}
\usepackage{pos}

\title{Progress on Constraining the Strange Quark Contribution to the Nucleon Spin}

\author*[a]{S.F. Pate}
\author[a]{V. Papavassiliou}
\author[a]{J.P. Schaub}
\author[a]{D.P. Trujillo}
\author[b]{M.V. Ivanov}
\author[c,d]{M.B. Barbaro}
\author[e]{C. Giusti}

\affiliation[a]{Department of Physics, New Mexico State University, Las Cruces, NM, USA}
\affiliation[b]{Institute for Nuclear Research and Nuclear Energy, Bulgarian Academy of Sciences, Sofia 1784, Bulgaria}
\affiliation[c]{Universit\`a degli Studi di Torino, Turin, Italy}
\affiliation[d]{Istituto Nazionale di Fisica Nucleare, Sezione di Torino, Italy}
\affiliation[e]{Istituto Nazionale di Fisica Nucleare, Sezione di Pavia, Italy}

\emailAdd{spate@nmsu.edu}

\abstract{We report on a global fit of neutral-current elastic (NCE) neutrino-scattering data and parity-violating electron-scattering (PVES) data with the goal of determining the strange quark contribution to the vector and axial form factors of the proton. Knowledge of the strangeness contribution to the axial form factor, $G_A^s(Q^2)$, at low $Q^2$ will reveal the strange quark contribution to the nucleon spin, as $G_A^s(Q^2=0)=\Delta s$. Previous fits~\cite{Pate:2008va,Pate:2013wra} of this form included data from a variety of PVES experiments (PVA4, HAPPEx, G0, SAMPLE) and the NCE neutrino and anti-neutrino data from BNL E734. These fits did not constrain $G_A^s(Q^2)$ at low $Q^2$ very well because there was no NCE data for $Q^2<0.45$ GeV$^2$. Our new fit includes for the first time MiniBooNE NCE data from both neutrino and anti-neutrino scattering; this experiment used a hydrocarbon target and so a model of the neutrino interaction with the carbon nucleus was required. Three different nuclear models have been employed; a relativistic Fermi gas (RFG) model, the SuperScaling Approximation (SuSA) model, and a spectral function (SF) model~\cite{Giusti:2019cup}. We find a tremendous improvement in the constraint of $G_A^s(Q^2)$ at low $Q^2$ compared to previous work, although more data is needed from NCE measurements that focus on exclusive single-proton final states, for example from MicroBooNE~\cite{Ren:2022qop}. This work has been published in Physical Review D~\cite{PhysRevD.109.093001}.}

\FullConference{
The 26th international symposium on spin physics (SPIN2025)\\
21–26 September, 2025\\
Qingdao (Tsingtao), Shandong Province, China\\}

\begin{document}
\maketitle

\section{Physics Motivation}

The axial form factor of the proton $G_A(Q^2)$ is the leading contributor to the interactions of neutrinos with matter, just as the electric and magnetic form factors are the leading contributors to the interactions of electrons with matter.   It may be written as a sum over contributions from individual quarks.
$$G_A(Q^2) = \frac{1}{2}\left[-G_A^u(Q^2)+G_A^d(Q^2)+G_A^s(Q^2)\right] $$
The up-down part,
$$G_A^{CC}(Q^2) = -G_A^u(Q^2)+G_A^d(Q^2) $$
is very well-known from decades of study of charged-current (CC) interactions, for example $\nu_\mu+n\rightarrow\mu+p$ and $n\rightarrow p+e^-+\bar{\nu}_e$.  The strange part,
$$G_A^s(Q^2)$$
by contrast is only directly accessible via neutral-current (NC) interactions, for example $\nu+p\rightarrow\nu+p$, for which there is very little data compared to CC interactions.  We have still only limited information about $G_A^s(Q^2)$.

In practical models of neutrino interactions, it is necessary to create some form for $G_A^s(Q^2)$.  Two common ingredients are found in these models.
\begin{itemize}
\item It is assumed that the $Q^2$-dependence of $G_A^s(Q^2)$ is the same as $G_A^{CC}(Q^2)$, that is a dipole form using an ``axial mass'' $M_A$ taken from studies of CC data.  But there is no physics underlying this assumption.
\item The value of $G_A^s(Q^2)$ at $Q^2=0$ is the strange quark contribution to the proton spin, usually called $\Delta s$.  A value for $\Delta s$ from polarized deep-inelastic scattering data is then taken for use in the model.  However, there is no agreed upon value for $\Delta s$ from pDIS data; values range anywhere from 0.0 to -0.2.  This results in a big uncertainty in the modeling.
\end{itemize}
Our goal here is to determine both the $Q^2$-dependence of $G_A^s(Q^2)$ and the value of $\Delta s$ directly from elastic electron and neutrino scattering data.

\begin{figure}[ht]
\centering
\includegraphics[scale=0.5]{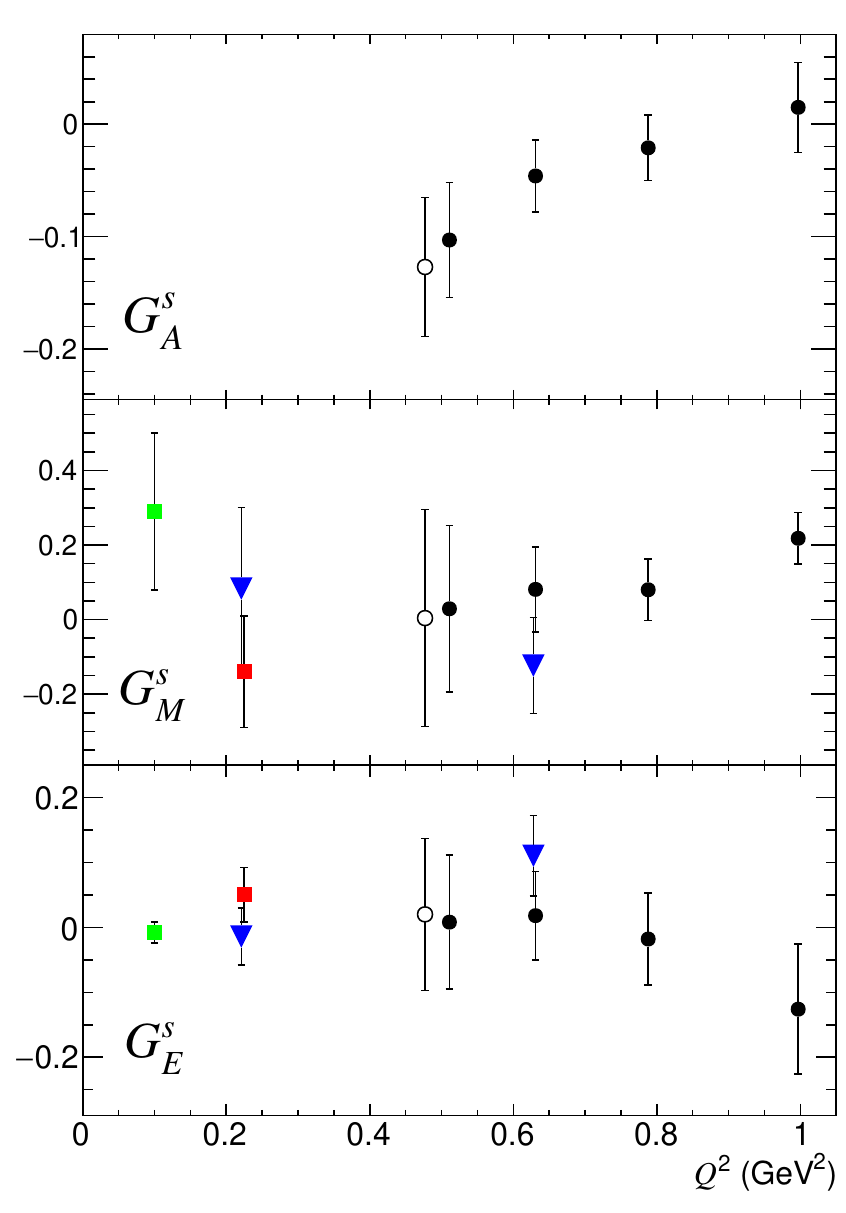}
\caption{Independent determinations of the strangeness form factors of the nucleon using subsets of existing experimental data: Liu et al.\ (green squares) \cite{Liu:2007yi}; Androi\'{c} et al.\ (blue triangles) \cite{Androic:2009zu}; Baunack et al.\ (red squares) \cite{Baunack:2009gy}; Pate et al.\ (open circles use HAPPEx and E734 data, and closed circles use G0-Forward and E734 data) \cite{Pate:2008va}.  This selection of results is representative and not intended to be exhaustive.}
\label{fig:old_points}
\end{figure}

Our approach will be to determine simultaneously the strange quark contribution to the electric, magnetic and axial form factors 
($G_E^s(Q^2)$, $G_M^s(Q^2)$, and $G_A^s(Q^2)$ respectively)
by combining data from neutrino neutral current
elastic scattering (NCES) and parity-violating
electron scattering (PVES).
This was first done in \cite{Pate:2003rk} by
combining BNL E734 NCES data with HAPPEx PVES
data at $Q^2 = 0.477$ GeV$^2$.
This analysis was expanded \cite{Pate:2008va}
to include points in the range $0.55 < Q^2 < 1.05$ GeV$^2$
when the G0 PVES data became available.  Other analyses focused only on the PVES data and extracted
only $G_E^s(Q^2)$ and $G_M^s(Q^2)$.   The results of these various analyses are summarized in Figure~\ref{fig:old_points}.
We observe that the strange quark contribution to the electric and magnetic form factors, $G_E^s(Q^2)$ and $G_M^s(Q^2)$, are consistent with zero throughout the range $0.0 < Q^2 < 1.1$ GeV$^2$.  On the other hand, $G_A^s(Q^2)$ has a definite $Q^2$-dependence, trending negative with decreasing $Q^2$.

\section{Description of the Method}

To make progress with the determination of these form factors, we need to transition from determining the form factors at local values of $Q^2$ to a global fit that uses all of the data points.  This means we need parametrized functional forms for $G_E^s(Q^2)$, $G_M^s(Q^2)$, and $G_A^s(Q^2)$.
Based on the results seen in Figure~\ref{fig:old_points}, we have chosen very simple, zeroth-order forms for $G_E^s(Q^2)$ and $G_M^s(Q^2)$.
$$G_E^s(Q^2) = \rho_s \tau ~~~~~~~~~~ \tau = Q^2/4M^2$$
$$G_M^s(Q^2) = \mu_s$$
where $\rho_s$ is the strangeness radius, and $\mu_s$ is the strangeness contribution to the magnetic moment.  
By contrast, the data on $G_A^s$ shows a definite $Q^2$-dependence, and for this form factor we have chosen to use two different 3-parameter models.
\begin{itemize}
\item The Modified-Dipole Model:  The expression used for the strangeness axial form factor is
$$G_A^s = \frac{\Delta s + S_A Q^2}{(1+Q^2/\Lambda_A^2)^2}$$
where $\Delta s$ is the strange quark contribution to the proton spin, and $S_A$ and $\Lambda_A$ are parameters describing the $Q^2$-dependence of $G_A^s$.
This shape is referred to as a ``modified-dipole'' because of its similarity to the usual dipole shapes used to model other form factors.
\item The $z$-Expansion Model:  The modified-dipole model comes with a bias with respect to the $Q^2$-dependence of $G_A^s$.  The ``$z$-expansion'' technique \cite{Hill:2010yb,Lee:2015jqa} allows for a bias-free model because it is simply a power series, and the fit seeks to determine the coefficients of the series.  The power series is of the form
$$G_A^s(Q^2)=\sum_{k=0}^{k_{\rm max}} a_k\left[z(Q^2)\right]^k
~~~~~~~~ z(Q^2)=\frac{\sqrt{(4 m_\pi)^2+Q^2}-\sqrt{(4 m_\pi)^2}}{\sqrt{(4 m_\pi)^2+Q^2}+\sqrt{(4 m_\pi)^2}}.$$
Note that $|z|<1$.   We have limited the sum to $k_{\rm max}=6$.  This would imply seven 
parameters for the description of $G_A^s$.  However, due to the fact that the form factor should behave 
like $1/Q^4$ at large values of $Q^2$, we have the following four conditions:
$$\left.\frac{d^n}{dz^n}G_A^s\right|_{z=1}=0~~~~~n=0,1,2,3.$$
This allows us to reduce the number of independent parameters from seven down to three:  $a_0$, $a_1$, and $a_2$.
\end{itemize}

A wide range of neutrino-scattering and electron-scattering are used in our analysis, from the E734, G0, PVA4, SAMPLE, HAPPEx and MiniBooNE experiments, comprising 128 data points.  In order to incorporate the MiniBooNE experiment, which used a hydrocarbon-based target/detector system, a model was needed for the neutrino-carbon interaction.  We used three such models.
In the Relativistic Fermi Gas (RFG), the carbon nucleus is described by a Fermi momentum $k_F$ based on electron scattering data; 
the nucleons are plane waves constrained by the Pauli principle;
in the SuperScaling Approximation (SuSA) model, scaling behavior of $(e,e')$ data are used to predict NC and CC neutrino-scattering cross sections; and
in the Spectral Function (SF) model, a spectral function $S(p,{\cal E})$ based on $(e,e')$ data has been used to better describe single-nucleon removal.

\begin{figure}[ht]
\centering
\includegraphics[scale=0.5]{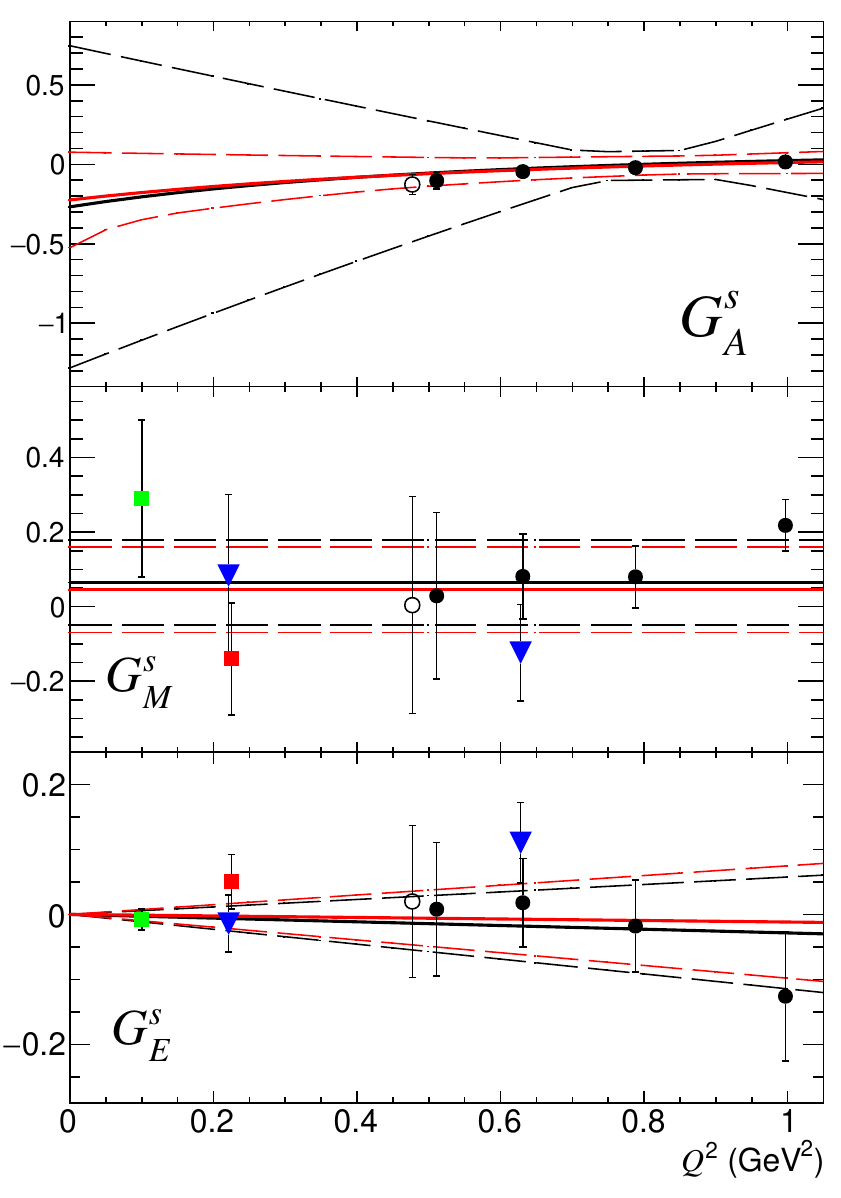}
\caption{An illustration of the effect of the introduction of the MiniBooNE neutral current data into our global fit.  The data points are the same as in Fig.~\ref{fig:old_points}.  The black solid line is the central value for the modified-dipole fit not using the MiniBooNE data. The red solid line includes the MiniBooNE data using the spectral function nuclear model.  The dashed lines represent the 70\% confidence limit for each fit. As mentioned in the text, the vector form factors fit is only slightly affected by the introduction of the MiniBooNE data, while the constraints on the axial form factor are greatly improved.}
\label{fig:reduction}
\end{figure}

Using the three nuclear models for carbon, and the two models for the strangeness form factors, we performed 6 distinct fits.  In each fit, the five form factor parameters were varied to find the minimum $\chi^2$, and the behavior of the $\chi^2$ near the minimum was used to determine the uncertainties in the parameters.  The complete results are described in great detail in our publication~\cite{PhysRevD.109.093001}.  In these proceedings we will simply illustrate in Figure~\ref{fig:reduction} the effect of adding the MiniBooNE data to this analysis.  It is seen that the constraint on the low-$Q^2$ behavior of $G_A^s(Q^2)$ is greatly improved by the introduction of the MiniBooNE data, while there is little change to the other form factors $G_E^s(Q^2)$ and $G_M^s(Q^2)$.

An important ingredient, which is missing in the present models, is the contribution of two-body currents, which can lead to the  excitation of 2p2h states. These contributions are not quasi-elastic, but they do contribute to the experimental signal and should be included in the calculation. In principle two-body currents could affect not only the cross sections but also the p/n ratios because of the isospin dependence of the current operator. However, while several calculations are now available for the 2p2h contribution to CC reactions, the corresponding calculations for NC scattering are very rare.

\section{Conclusion}

We have performed a global fit of parity-violating electron-scattering data from the HAPPEx, SAMPLE, G0 and PVA4 experiments and of neutral-current elastic scattering data from the BNL E734 and Fermilab MiniBooNE experiments, a total of 128 data points in the momentum transfer range $0.1 < Q^2 < 1.1$ GeV$^2$, using two models for the strangeness form factors $G_E^s$, $G_M^s$, and $G_A^s$, and using three nuclear models to describe the interaction of neutrinos with the hydrocarbon target used in MiniBooNE. 
Our fits are in very good agreement with this collection of data, with $\chi^2$/ndf $\approx$ 1.1-1.2 for all fits.  

Depending on the model, we show a slightly negative value of the strangeness radius $\rho_s$ but also consistent with zero, and a slightly positive value for the strangeness magnetic moment $\mu_s$ also consistent with zero.  To quantify our conclusion that $\rho_s$ and $\mu_s$ are consistent with zero, we have taken $\rho_s=0$ and $\mu_s=0$ to be null hypotheses and then used our fit results for these quantities to calculate a corresponding p-value for each.  For the null hypothesis $\rho_s=0$ we find a p-value of 0.83; for the null hypothesis $\mu_s=0$ we find a p-value of 0.42.  These large p-values do not recommend a rejection of either of these null hypotheses.

The inclusion of the MiniBooNE neutral current data into the dataset has greatly improved the constraints on the strangeness axial form factor $G_A^s$, but still we cannot report a definite value for $\Delta s$ on the basis of these fits.
We can expect that a more refined model including two-body currents (which is currently not available but can hopefully become available in the future) would give a better description of the experimental NC cross section and might be helpful for an improved determination of the strange axial form factor, but presumably it should not change the main finding of our paper that the inclusion of the MiniBooNE neutral current data into the dataset greatly improves the constraints on $G_A^s$. 
Primarily, exclusive NCES data from proton interactions at low $Q^2$ are still needed for a complete determination of $G_A^s$, and we look forward to that data from MicroBooNE~\cite{Ren:2022qop} in the near future.

\section*{Acknowledgements}
We are grateful to the following funding agencies:  US Department of Energy, Office of Science, Medium Energy Nuclear Physics Program, Grant DE-FG02-94ER40847; Istituto Nazionale di Fisica Nucleare under the National Project ``NUCSYS''; University of Turin local research funds BARM-RILO-22.
SFP is grateful for sabbatical support from both the Universities Research Association Visiting Scholars Program and the Los Alamos National Laboratory during 2022.
We are also grateful to R.~Dharmapalan for assistance in the interpretation of the MiniBooNE antineutrino NC data release.

\bibliographystyle{unsrt}
\bibliography{pate_spin2025}

\end{document}